\documentclass{egpubl}
\usepackage{pg2023p}

\WsPoster          % uncomment for Poster contribution
\usepackage[T1]{fontenc}
\usepackage{dfadobe}  

\usepackage{cite}  % comment out for biblatex with backend=biber
\BibtexOrBiblatex
\electronicVersion
\PrintedOrElectronic
\ifpdf \usepackage[pdftex]{graphicx} \pdfcompresslevel=9
\else \usepackage[dvips]{graphicx} \fi

\usepackage{egweblnk} 
\title[Hand Shadow Art: A Differentiable Rendering Perspective]%
      {Hand Shadow Art: A Differentiable Rendering Perspective \vspace{-0.5cm}}

\author[A. Gangopadhyay, P. Singh, A. Tiwari \& S. Raman]{Aalok Gangopadhyay, Prajwal Singh, Ashish Tiwari \& Shanmuganathan Raman \\ CVIG Lab, Indian Institute of Technology Gandhinagar \vspace{-0.3cm}}
\begin{document}

% % uncomment for using teaser
% \teaser{
%  \includegraphics[width=\linewidth]{images/approach.png}
%  \centering
%   \caption{Caption}
% \label{fig:teaser}
% }
\maketitle
%-------------------------------------------------------------------------
\begin{abstract}
   Shadow art is an exciting form of sculptural art that produces captivating artistic effects through the 2D shadows cast by 3D shapes. Hand shadows, also known as shadow puppetry or shadowgraphy, involve creating various shapes and figures using your hands and fingers to cast meaningful shadows on a wall. In this work, we propose a differentiable rendering-based approach to deform hand models such that they cast a shadow consistent with a desired target image and the associated lighting configuration. We showcase the results of shadows cast by a pair of two hands and the interpolation of hand poses between two desired shadow images. We believe that this work will be a useful tool for the graphics community.
%-------------------------------------------------------------------------
%  ACM CCS 1998
%  (see https://www.acm.org/publications/computing-classification-system/1998)
% \begin{classification} % according to https://www.acm.org/publications/computing-classification-system/1998
% \CCScat{Computer Graphics}{I.3.3}{Picture/Image Generation}{Line and curve generation}
% \end{classification}
%-------------------------------------------------------------------------
%  ACM CCS 2012
   % (see https://www.acm.org/publications/class-2012)
%The tool at \url{http://dl.acm.org/ccs.cfm} can be used to generate
% CCS codes.
%Example:


\end{abstract}  
%-------------------------------------------------------------------------
\section{Introduction}
\textit{"Will he not fancy that the shadows which he formerly saw are truer than the objects which are now shown to him?"} - Plato, the Republic.\\
\indent Shadows are crucial in how we perceive the world around us. They are the silent storytellers of our world, revealing secrets that light alone cannot. For a long time, it has been gathering the attention of artists, stage performers, and filmmakers. Hand shadows and other forms of storytelling, such as ancient Chinese shadow puppetry, have been instrumental in unleashing the capacity of the human brain to recognize 3D objects through shadows. Hand shadows offer a versatile and imaginative form of shadow art that can be used for entertainment, education, artistic expression, and various cultural and creative expression forms. In this work, we attempt to answer how we can deform our hands such that the shadow thus cast matches the desired image using differentiable rendering \cite{ravi2020pytorch3d} (see Figure \ref{fig:bd}).\\
\indent While a few recent methods have addressed shadow art using either optimization \cite{mitra2009shadow} or differentiable rendering \cite{sadekar2022shadow}, none of them have addressed hand shadow art. Mitra \emph{et al.} \cite{mitra2009shadow} described shadow art more formally by introducing a voxel-based optimization framework to recover the 3D shape from arbitrary input (shadow) images by deforming them and handling inherent image inconsistencies. Sadekar \emph{et al.} \cite{sadekar2022shadow} demonstrated the potential of differentiable rendering (mesh and voxel-based) in generating 3D shadow sculptures all from arbitrary shadow images without any explicit input image deformation. In this work, we explicitly focus on developing a differential rendering-based optimization framework to create hand shadow art by deforming the hand mesh models and show the dynamics involved in interpolating between two shadow images. To the best of our knowledge, this is the first such attempt to generate hand shadow art progressively using differentiable rendering.

\begin{figure}[t]
  \centering
  \includegraphics[width=\linewidth]{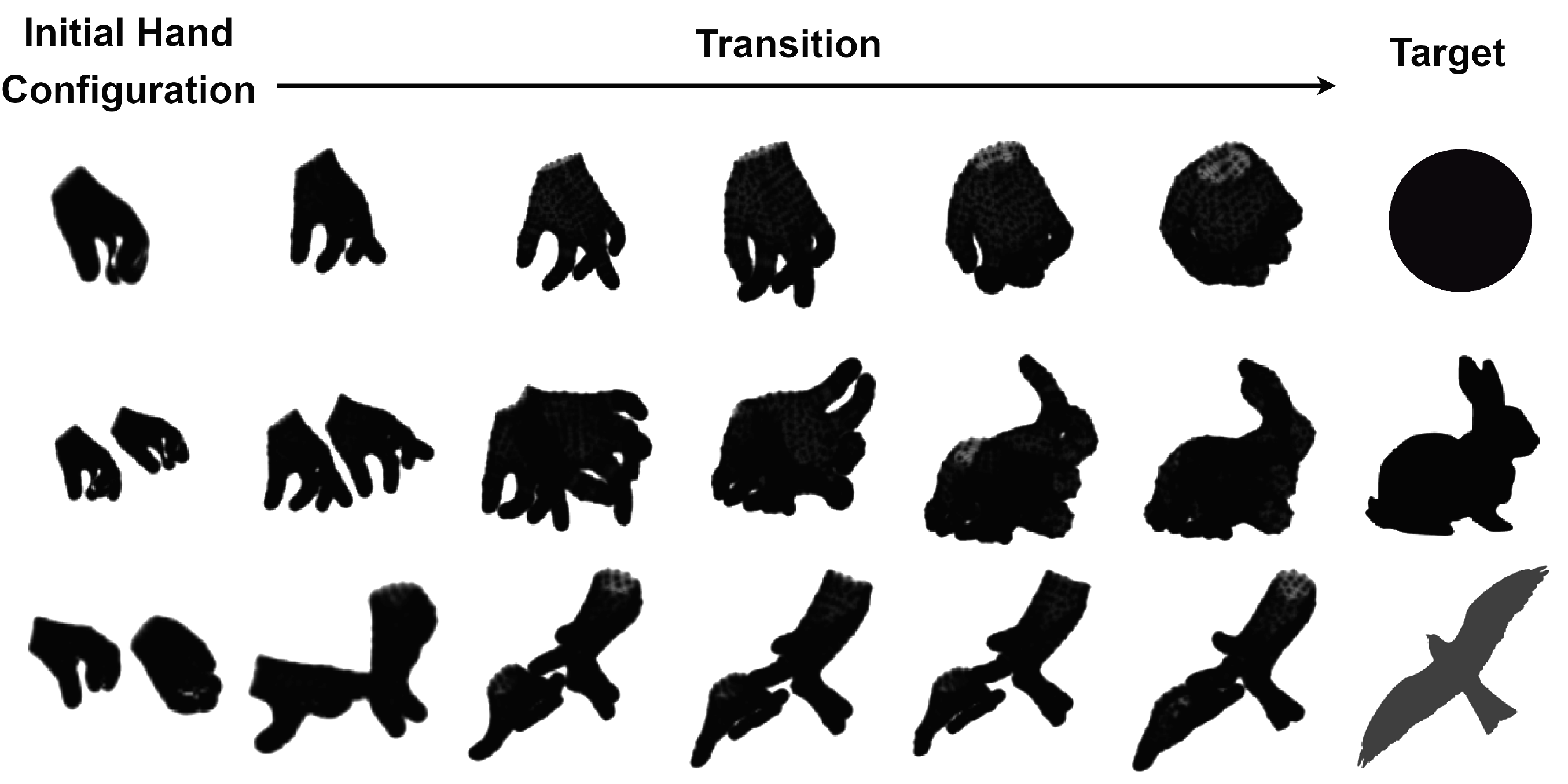}
  \caption{Different hand configurations to render the target shadow image during the optimization process from one hand (first row) and two hands (second and third row). The images are rendered in $256 \times 256$ resolution.}
  \label{fig:res_1}
\vspace{-0.35cm}
\end{figure}

\section{Method}

\begin{figure}[t]
  \centering
  \includegraphics[width=\linewidth]{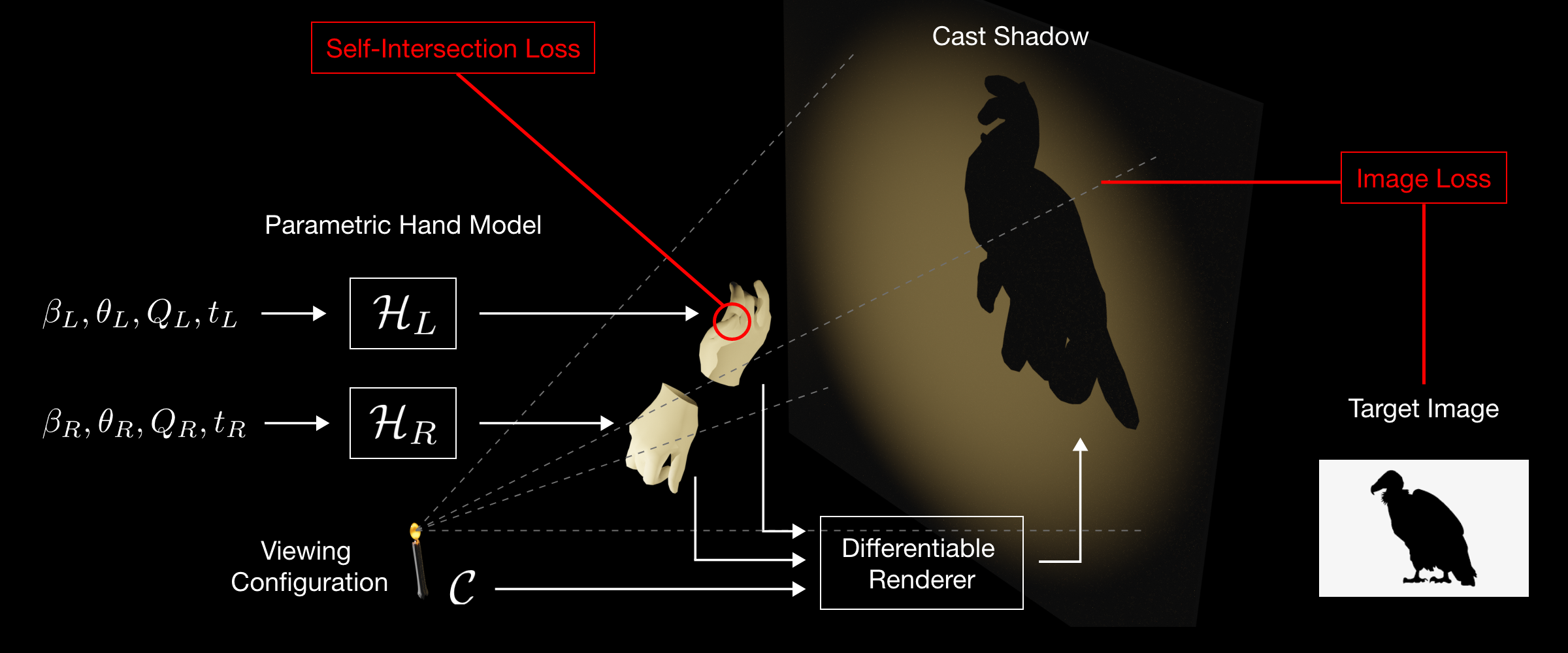}
  \caption{The outline of the proposed differentiable rendering framework to simulate hand shadow art. The framework is optimized over $5000$ iterations to reach convergence.}
  \label{fig:bd}
\end{figure}

The design of the proposed framework is described in Figure \ref{fig:bd}. Let $\mathcal{I}^{*} = [0,1]^{H \times W}$ denote the space of all grayscale images defined on a grid of size $H \times W$. Let $I \in \mathcal{I}^{*}$ be a given target image and $\mathcal{M} = \mathcal{H}(\beta, \theta, Q, t)$ be the parametric hand model \cite{romero2022embodied}. Here, $\beta$, $\theta$, $Q$, and $t$ are the shape parameter, pose parameter, rotation matrix, and translation vector, with $\mathcal{M}$ being the mesh model of the hand for a given set of parameters. Please note that we use the parametric models for both the left and right hand given by $\mathcal{M}_{L} = \mathcal{H}_L(\beta_L, \theta_L, Q_L, t_L)$ and $\mathcal{M}_{R} = \mathcal{H}_R(\beta_R, \theta_R, Q_R, t_R)$. Consider $\mathcal{C}$ as the camera with fixed parameters placed at the origin. The lighting and camera parameters together are termed as \textit{viewing configuration} $(\mathcal{V})$. Let $\mathcal{R}(\mathcal{C},\mathcal{M}_{L},\mathcal{M}_{R})$ denote the silhouette image rendering of both the hands as seen from the camera. The objective is to find the values of parameters $\theta_L$, $\theta_R$, $Q_L$, $Q_R$, $t_L$, and $t_R$ that minimize $||I - \mathcal{R}(\mathcal{C},\mathcal{M}_L,\mathcal{M}_R)||_{2}$, keeping the shape parameters ($\beta_L$ and $\beta_R$) fixed.  The non-rigid nature of hands and fingers poses an additional challenge apart from rigid transformations \emph{i.e.} rotations and translations and mesh intersections towards learning optimal hand configurations. To avoid self-intersections and cross-intersections among a pair of hand meshes, we penalize such intersections according to pen loss \cite{Karras:2012:MPC:2383795.2383801, Tzionas:IJCV:2016}. Further, we restrict the angular movement across $15$ different joints per hand in the MANO hand models \cite{romero2022embodied} to simulate realistic and plausible human hand movements (see Table \ref{tab:1}).

\begin{figure}[t]
  \centering
  \includegraphics[width=\linewidth]{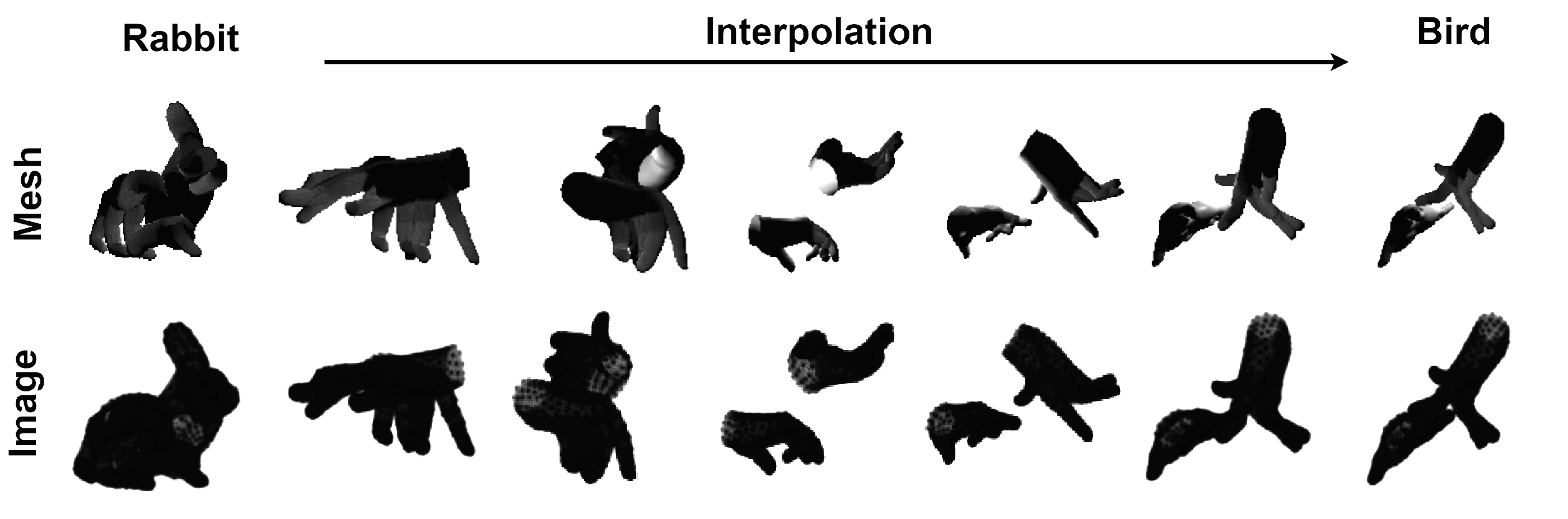}
  \caption{Different hand configurations (top: hand meshes, bottom: rendered image) while interpolating from a Rabbit to a Bird silhouette. Here is a \href{https://gifyu.com/image/SgUsb}{link to a dynamic visualization}.}
  \label{fig:interp}
\end{figure}

\section{Results and Discussion}
Figure \ref{fig:res_1} shows a few transitional hand configurations to cast the shadow resembling a target image and also transitions between one image to another.

\textbf{Hand shadow art.} We start with a random initial configuration of hand models to reach the final pose, creating a shadow similar to the given target image when viewed from a fixed configuration. 

\textbf{Interpolation} between a pair of shadow images cast by a pair of hands is shown in Figure \ref{fig:interp}. Given a pair of images $I_A$ and $I_B$, we find a sequence of hand mesh models$ (\mathcal{M}_{L}^{t}, \mathcal{M}_{R}^{t})$ such that $\mathcal{R}(\mathcal{C}, \mathcal{M}_{L}^{0}, \mathcal{M}_{R}^{0})=I_{A}$ and $\mathcal{R}(\mathcal{C},\mathcal{M}_{L}^{T}, \mathcal{M}_{R}^{T})=I_{B}$, where $t \in [0,T]$.

\begin{figure}[t]
  \centering
  \includegraphics[width=\linewidth]{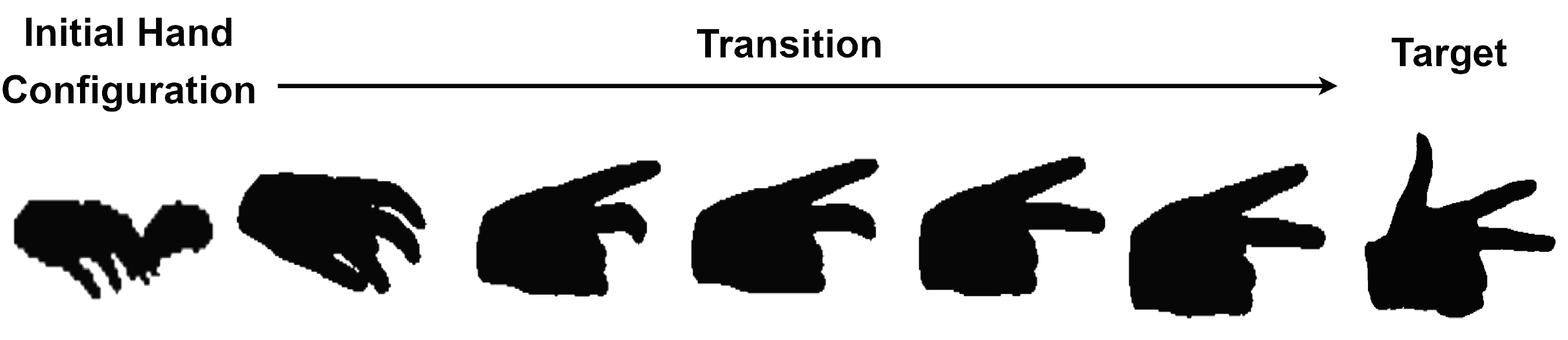}
  \caption{Failure cases over large and abrupt transitions and sensitivity to the initial hand configuration.}
  \label{fig:lim}
\end{figure}

We observe that a good initial hand configuration makes the convergence faster and easier, otherwise it leads to incorrect results. For example, the model cannot capture a larger transition of the thumb to recreate the target hand configuration from a specific initial configuration (see Figure \ref{fig:lim}). In the near future, we wish to work towards adapting the algorithm to be free of specific initialization and extend the current single-view setting to a multi-view setting where a single-hand configuration can cast different meaningful shadows when viewed and lit from multiple directions.

\begin{table}[t]
\setlength{\tabcolsep}{0.1pt}
\centering
\resizebox{\linewidth}{!}{
\begin{tabular}{ccccccccccc} \hline
\begin{tabular}[c]{@{}c@{}}Joints \\ Name/\\ Locations\end{tabular} &
  \begin{tabular}[c]{@{}c@{}}RH\\ Index\end{tabular} &
  \begin{tabular}[c]{@{}c@{}}RH\\ Middle\end{tabular} &
  \begin{tabular}[c]{@{}c@{}}RH\\ Little\end{tabular} &
  \begin{tabular}[c]{@{}c@{}}RH\\ Ring\end{tabular} &
  \begin{tabular}[c]{@{}c@{}}RH\\ Thumb\end{tabular} &
  \begin{tabular}[c]{@{}c@{}}LH\\ Index\end{tabular} &
  \begin{tabular}[c]{@{}c@{}}LH\\ Middle\end{tabular} &
  \begin{tabular}[c]{@{}c@{}}LH\\ Little\end{tabular} &
  \begin{tabular}[c]{@{}c@{}}LH\\ Ring\end{tabular} &
  \begin{tabular}[c]{@{}c@{}}LH\\ Thumb\end{tabular} \\ \hline
J.1 $(\theta, \phi, \psi)$ &
  \begin{tabular}[c]{@{}c@{}}0.0,\\ -15.0,\\ -35.0\end{tabular} &
  \begin{tabular}[c]{@{}c@{}}0.0,\\ -25.0,\\ -15.0\end{tabular} &
  \begin{tabular}[c]{@{}c@{}}0.0,\\ -40.0,\\ -35.0\end{tabular} &
  \begin{tabular}[c]{@{}c@{}}0.0,\\ -15.0,\\ -35.0\end{tabular} &
  \begin{tabular}[c]{@{}c@{}}-50.0,\\ -10.0,\\ -30.0\end{tabular} &
  \begin{tabular}[c]{@{}c@{}}0.0,\\ 25.0,\\ 75.0\end{tabular} &
  \begin{tabular}[c]{@{}c@{}}0.0,\\ 10.0,\\ 75.0\end{tabular} &
  \begin{tabular}[c]{@{}c@{}}0.0,\\ 25.0,\\ 75.0\end{tabular} &
  \begin{tabular}[c]{@{}c@{}}0.0,\\ 15.0,\\ 75.0\end{tabular} &
  \begin{tabular}[c]{@{}c@{}}20.0,\\ 30.0,\\ 40.0\end{tabular} \\
J.2 $(\theta, \phi, \psi)$ &
  \begin{tabular}[c]{@{}c@{}}0.0,\\ 0.0,\\ -45.0\end{tabular} &
  \begin{tabular}[c]{@{}c@{}}0.0,\\ 0.0,\\ -15.0\end{tabular} &
  \begin{tabular}[c]{@{}c@{}}0.0,\\ 0.0,\\ -45.0\end{tabular} &
  \begin{tabular}[c]{@{}c@{}}0.0,\\ 0.0,\\ -45.0\end{tabular} &
  \begin{tabular}[c]{@{}c@{}}-10.0,\\ -20.0,\\ -10.0\end{tabular} &
  \begin{tabular}[c]{@{}c@{}}0.0,\\ 0.0,\\ 45.0\end{tabular} &
  \begin{tabular}[c]{@{}c@{}}0.0,\\ 0.0,\\ 55.0\end{tabular} &
  \begin{tabular}[c]{@{}c@{}}0.0,\\ 0.0,\\ 55.0\end{tabular} &
  \begin{tabular}[c]{@{}c@{}}0.0,\\ 0.0,\\ 55.0\end{tabular} &
  \begin{tabular}[c]{@{}c@{}}10.0,\\ 20.0,\\ 20.0\end{tabular} \\
J.3 $(\theta, \phi, \psi)$ &
  \begin{tabular}[c]{@{}c@{}}0.0,\\ 0.0,\\ -15.0\end{tabular} &
  \begin{tabular}[c]{@{}c@{}}0.0,\\ 0.0,\\ -45.0\end{tabular} &
  \begin{tabular}[c]{@{}c@{}}0.0,\\ 0.0,\\ -5.0\end{tabular} &
  \begin{tabular}[c]{@{}c@{}}0.0,\\ 0.0,\\ -15.0\end{tabular} &
  \begin{tabular}[c]{@{}c@{}}-5.0,\\ -50.0,\\ 0.0\end{tabular} &
  \begin{tabular}[c]{@{}c@{}}0.0,\\ 0.0,\\ 75.0\end{tabular} &
  \begin{tabular}[c]{@{}c@{}}0.0,\\ 0.0,\\ 75.0\end{tabular} &
  \begin{tabular}[c]{@{}c@{}}0.0,\\ 0.0,\\ 65.0\end{tabular} &
  \begin{tabular}[c]{@{}c@{}}0.0,\\ 0.0,\\ 75.0\end{tabular} &
  \begin{tabular}[c]{@{}c@{}}5.0,\\ 25.0\\ 0.0\end{tabular} \\ \hline
\end{tabular}}
\caption{$(\theta, \phi, \psi)$ represents rotation about $(x,y,z)$ axis and J.1, J.2, J.3 represents joints along each finger of the left (LH) and right (RH) hand.}
\label{tab:1}
\end{table}

\section{Conclusion}
In this work, we used a differentiable rendering-based approach to simulate hand shadow art through optimization. While we could create a variety of shadows through hand models, including interpolation, we observed that the framework is sensitive to initial hand configuration and susceptible to errors under large and abrupt transitions, which we aim to address in the near future, including dynamic shadow art. This versatile medium can be applied to various contexts, from traditional storytelling and entertainment to contemporary art installations and experimental exhibitions, and we hope it will be helpful for the graphics community.

 % Motivate Hand shadow Art - When there is a light source in the scene, we can place our hands in front of the light source and deform our hands so that the shadow cast on the nearby wall forms a meaningful image. However given a desired image, how do we deform our hands so that the shadow cast matches the desired image? This brings us to the problem of hand shadow art which we formally define below.

% 

% \section{To-Do}

% 1) Cite the work for avoiding mesh intersection
% 2) Limits imposed on $\theta_L$ and $\theta_R$ for realistic hand deformation.
% 3) Our own customized differentiable renderer in constrast to pytorch silhouette renderer.

\vspace{-0.25cm}
\bibliographystyle{eg-alpha-doi} 
\small{\bibliography{egbibsample} }

\end{document}